\documentclass[showpacs,preprintnumbers,amsmath,amssymb]{revtex4}

\usepackage{graphicx}
\usepackage{subfigure}
\usepackage{dcolumn}
\usepackage{bm}

\begin{document}


\title{Derivation of quantum master equation with counting fields by monitoring a probe}

\date{\today}
\author{Takaaki Monnai}
\email{monnai@a-phys.eng.osaka-cu.ac.jp}%
\affiliation{$*$Department of Applied Physics, Osaka City University,
3-3-138 Sugimoto, Sumiyoshi-ku, Osaka 558-8585, Japan}
\begin{abstract}
We show a microscopic derivation of a quantum master equation with counting terms which describes the electron statistics.
A localized spin behaves as a probe whose precession angle monitors the net electron current by the magnetic-moment interaction.
The probe Hamiltonian is proportional to the current, and is determined self-consistently for a model of a quantum dot.
Then it turns out that the quantum master equation for the spin-precession contains the counting terms.
As an application, we show the fluctuation theorem for the electron current.    
\end{abstract}
\maketitle
\section{Introduction}
Recently the method of generating function has been extended to count the particle number flowing in quantum junctions
based on the quantum master equation with counting fields\cite{Esposito,Esposito2}. 
Indeed, the scheme has been successfully applied to the electron transport in quantum dots\cite{Esposito,Esposito2}, which provides a kinetic description of the recent experiments such as bidirectional electron transport in double quantum dots\cite{Utsumi,Fujisawa}.
Then the electron counting statistics illustrates the stochastic trajectory\cite{Harbola,Mukamel}.   
However, it remains unclear how the counting terms are derived microscopically from the interaction between the current and probe, i.e. we should give physical justifications of counting fields in the quantum master equation.
The problem here is to derive the quantum master equation with counting terms from the total Hamiltonian including the probe.
Although we only concern with the dynamics of the probe, the elimination of the system degrees of freedom is not self-evident due to the quantum correlation between the system and probe\cite{Levitov,Utsumi}.
Also, inclusion of the probe amounts to a self consisitent determination of the total Hamiltonian as shown in Sec.II.

In this article, we address this issue based on a spin probe for the electron transport\cite{Levitov}, and give a physical justification of counting terms which appear in the quantum master equation.
As an application, we show that the electron current satisfies a universal symmetry of fluctuation theorem\cite{Esposito,Esposito2,Utsumi,Esposito3,Harbola,Monnai,Andrieux,Utsumi2,Lebowitz}.

This paper is organized as follows.
In Sec.II, we present a model of electron transport and an idealistic spin probe which monitors the current.
In Sec.III, a quantum master equation with counting terms is derived.
In Sec.IV, we derive the fluctuation theorem for net electron transfer based on the quantum master equation.
\section{Model}
Let us consider the electron current between the left and right reservoirs\cite{Esposito,Esposito2,Levitov,Andrieux,Andrieux2,Utsumi2}.
As a model of a quantum dot, we consider a subsystem with a discrete energy level located between the reservoirs\cite{Esposito2}.
In order to detect the magnetic field caused by the electron current, we prepare a spin probe sufficiently near the 
current\cite{Levitov}.
The total Hamiltonian including the probe is given as
\begin{equation}
H=H_0+V-\frac{\lambda\hbar}{2}\sigma_zI \label{totalHamiltonian}
\end{equation}
where 
\begin{eqnarray}
&&H_0=\hbar\Omega a^+a+\sum_{j=1}^2 \int dk \hbar\omega_{kj}a_{kj}^+a_{kj} \nonumber \\
&&V=\sum_{j=1}^2 \int dk \hbar(u_{kj}a a_{kj}^++v_{kj}a_{kj}a^+) \nonumber \\
&&I=-\frac{1}{i\hbar}[\int dk a_{k1}^+a_{k1},H]. \label{current}
\end{eqnarray}
$H_0$ consists of Hamiltonians of the subsystem and reservoirs, where $a$, $a_{k1}$, and $a_{k2}$ are annihilation operators of the subsystem, left and right reservoirs, respectively.
In the case of electron transport, they satisfy the anti-commutation relations $\{a,a^+\}=1$, $\{a_{ki},a_{k'j}\}=\delta_{ij}\delta(k-k')$.
Also $V$ presents a bilinear coupling between the system and reservoirs, which yields the current.
According to the Ampere's law, the current yields magnetic field.
Then the spin-probe shows a precession whose precession angle is proportional to the current. 
The amplitude of the magnetic field is proportional to that of the current $I$, and the magnetic moment interaction energy is written as $-\frac{\lambda\hbar}{2}\sigma_zI$.
Here $\lambda$ is a measure of the coupling strength between the current and spin-probe, and $\sigma_z$ is $z$ component of Pauli matrices.  
The particle current $I$ is defined as the time derivative of the particle number of the left reservoir.
It is also possible to concern with the particle number of the right reservoir
\begin{equation}
I'=\frac{1}{i\hbar}[\int dk a_{k2}^+a_{k2},H],
\end{equation}
 which is equal to $I$ in the stationary state.   
To concern only with the left reservoir means that the probe monitors the electron moving between the left reservoir and the subsystem. 

Note that the definition of the total Hamiltonian $H$ includes the current $I$, which depends on $H$.
This condition amounts to a self-consistent equation for $H$, which is solved in the following subsection.
\subsection{Self-consistent determination of the total Hamiltonian}
In this subsection, we construct the total Hamiltonian which is consistent with the definition of the current $I$ defined in Eq.(\ref{current}). 
As the lowest order evaluation of the current, let us define $I_0$ as 
\begin{equation}
I_0=-\frac{1}{i\hbar}[\int dk a_{k1}^+a_{k1},H_0+V].
\end{equation}
It is straightforward to verify that 
\begin{equation}
I_0=-\frac{1}{i\hbar}\int dk \hbar(u_{k1}aa_{k1}^+-v_{k1}a_{k1}a^+).
\end{equation}
Then the first order evaluation of the total Hamiltonian is 
\begin{equation}
H_1=H_0+V-\frac{\lambda\hbar}{2}\sigma_z I_0.
\end{equation}
Similarly, the next order evaluation of the  current is given as
\begin{eqnarray}
&&I_1=-\frac{1}{i\hbar}[\int dk a_{k1}^+a_{k1}, H_1] \nonumber \\
&=&I_0+\frac{\lambda\hbar}{2}\sigma_z\frac{1}{(i\hbar)^2}\int dk \hbar(u_{k1}aa_{k1}^++v_{k1}a_{k1}a^+).
\end{eqnarray}
And the second order expression of the total Hamiltonian is 
\begin{equation}
H_2=H_0+V-\frac{\lambda\hbar}{2}\sigma_z I_1.
\end{equation}
In this way, we can recursively define the $m$-th evaluation of the total Hamiltonian $H_m$ as
\begin{eqnarray}
&&H_m=H_0+V-\frac{\lambda\hbar}{2}\sigma_z I_{m-1} \nonumber \\
&&I_m=-\frac{1}{i\hbar}[\int dk a_{k1}^+a_{k1},H_m].
\end{eqnarray}
Then it turns out that $\lim_{n\rightarrow\infty}H_n$ satisfies the self-consistent equation
\begin{equation}
\lim_{n\rightarrow\infty}H_n=H_0+V+\frac{\lambda\hbar}{2}\sigma_z\frac{1}{i\hbar}[\int dk a_{k1}^+a_{k1},\lim_{n\rightarrow\infty}H_n].
\end{equation}
More explicitly, we have 
\begin{eqnarray}
&&\lim_{n\rightarrow\infty}H_n\nonumber \\
&=&H_0+\int dk \hbar(u_{k1}(\lambda)aa_{k1}^++v_{k_1}(\lambda)a_{k1}a^+)+\int dk \hbar(u_{k2}aa_{k2}^++v_{k2}a_{k2}a^+) \nonumber \\
&&u_{k1}(\lambda)=\frac{1}{1+\frac{i\lambda}{2}\sigma_z}u_{k1} \nonumber \\
&&v_{k1}(\lambda)=\frac{1}{1-\frac{i\lambda}{2}\sigma_z}v_{k1}.
\end{eqnarray}
\subsection{Time evolution of the probe}
Depending on the value of $\sigma_z$, we use the abbreviated notation 
\begin{eqnarray}
&&H_\lambda=\langle\uparrow|\lim_{n\rightarrow\infty}H_n|\uparrow\rangle; \nonumber \\
&&H_{-\lambda}=\langle\downarrow|\lim_{n\rightarrow\infty}H_n|\downarrow \rangle.
\end{eqnarray}
Suppose that the initial state of the total system is described by the $2\times2$ density matrix 
\begin{equation}
\rho(0)=
\left(
\begin{array}{cc}
\rho_{\uparrow\uparrow} & \rho_{\uparrow\downarrow} \\
\rho_{\downarrow\uparrow} & \rho_{\downarrow\downarrow}
\end{array}
\right).
\end{equation}
Then the unitary time evolution is described as
\begin{equation}
\rho(t)=
\left(
\begin{array}{cc}
e^{-\frac{i}{\hbar}H_\lambda t}\rho_{\uparrow\uparrow}(0)e^{\frac{i}{\hbar}H_\lambda t} & e^{-\frac{i}{\hbar}H_\lambda t}\rho_{\uparrow\downarrow}(0)e^{\frac{i}{\hbar}H_{-\lambda}t} \\
e^{-\frac{i}{\hbar}H_{-\lambda}t}\rho_{\downarrow\uparrow}(0)e^{\frac{i}{\hbar}H_{\lambda t}} & e^{-\frac{i}{\hbar}H_{-\lambda}t}\rho_{\downarrow\downarrow}(0)e^{\frac{i}{\hbar}H_\lambda t}
\end{array}
\right).
\end{equation}
Since we only concern with the precession angle of the probe, the trace is taken for the subsystem and reservoirs variables.
Then the diagonal elements are invariant, while the off-diagonal elements evolves as
\begin{equation}
{\rm Tr_{s,r}}e^{-\frac{i}{\hbar}H_\lambda t}\rho_{\uparrow\downarrow}(0)e^{\frac{i}{\hbar}H_{-\lambda}t}. \label{spin}
\end{equation}
As pointed out in Ref.\cite{Levitov,Esposito3}, this quantity is identified as the characteristic function of the precession angle. 
In the following section, we pursue this issue in term of the corresponding quantum master equation.
\section{Quantum master equation}
In this section, we show our main result, i.e. a derivation of the quantum master equation with counting terms.   
Let us assume the weak coupling for $V$ between the subsystem and reservoirs\cite{Esposito,Esposito2}, and pursue the dynamics of the off-diagonal element of the density matrix.
As an initial condition, we assume that the left and right reservoirs are in mutually different equilibrium states and described by grand canonical ensembles, and the total density matrix is given as
\begin{equation}
\rho_{\uparrow\downarrow}(0)=\rho_s(0)\otimes \rho_r,
\end{equation}
where $\rho_s(0)$ is the density matrix of the subsystem, and $\rho_r$ is that of the reservoirs
\begin{equation}
\rho_r=\frac{1}{\Xi_1\Xi_2}e^{-\sum_{j=1}^2\beta_j\int dk (\hbar\omega_{kj}-\mu_j)a_{kj}^+a_{kj}}.
\end{equation}   
Here $\beta_j$ and $\mu_j$ are the inverse temperature and chemical potential, and $\Xi_j$ is the grand partition function of the $j$-th reservoir.
Then a quantum master equation for the system state ${\rm Tr_r}\rho$ is obtained in the Markovian limit:
In the second order perturbation with respect to $V$, the relaxation time of the subsystem $\tau_s$ is evaluated as 
\begin{equation}
\frac{1}{\tau_s}=O(\frac{u^2}{\hbar^2})\tau_c,
\end{equation}
where $u$ is a measure of the coupling strength between the subsystem and reservoirs, and $\tau_c$ is the typical correlation time of the reservoirs.   
In the weak coupling limit, and for sufficiently short $\tau_c$, we have 
\begin{equation}
\tau_s\gg\tau_c,
\end{equation}
 which justifies the Markovian description.

Let us derive the quantum master equation.
For this purpose, we use the interaction picture given as
\begin{eqnarray}
&&\rho_{\uparrow\downarrow}(t) \nonumber \\
&=&e^{-\frac{i}{\hbar}H_\lambda t}\rho_{\uparrow\downarrow}(0)e^{\frac{i}{\hbar}H_{-\lambda} t} \label{offdiagonal} 
\nonumber \\
&=&e^{-\frac{i}{\hbar}H_0 t}\sigma(t)e^{\frac{i}{\hbar}H_0 t}.
\end{eqnarray}
Then the von Neumann equation is written as 
\begin{equation}
\frac{\partial}{\partial t}\sigma(t)=\frac{1}{i\hbar}[e^{\frac{i}{\hbar}H_0 t}V(\lambda)e^{-\frac{i}{\hbar}H_0 t},\sigma(t)]_\lambda, \label{eq}
\end{equation}
where we have abbreviated the interaction Hamiltonian plus the probe Hamiltonian as
\begin{equation}
V(\lambda)=H_\lambda-H_0.
\end{equation}
Also the generalized commutator is defined as
\begin{equation}
[V(\lambda),A]_\lambda=V(\lambda)A-AV(-\lambda)
\end{equation}
for an arbitrary observable $A$.
We solve the equation as 
\begin{equation}
\sigma(t)=\sigma(0)+\frac{1}{i\hbar}\int_0^t ds[V(\lambda,s),\sigma(s)]_\lambda, \label{eq2}
\end{equation}
where $V(\lambda,s)=e^{\frac{i}{\hbar}H_0 s}V(\lambda)e^{-\frac{i}{\hbar}H_0 s}$.
By substituting Eq.(\ref{eq2}) into the right-hand-side of Eq.(\ref{eq}),
  von Neumann equation is expressed as 
\begin{eqnarray}
&&\frac{\partial}{\partial t}\sigma(t) \nonumber \\
&=&-\frac{1}{\hbar^2}\int_0^t ds[V(\lambda,t),[V(\lambda,s),\sigma(s)]_\lambda]_\lambda \nonumber \\
&&+\frac{1}{i\hbar}[V(\lambda,t),\sigma(0)]_\lambda. \label{eqmotion}
\end{eqnarray}
Note that 
\begin{equation}
V(\lambda,t)=\sum_j \int dk \hbar(u_{kj}(\lambda)aa_{kj}^+e^{i(\Omega-\omega_{kj})t}+v_{kj}(\lambda)a_{kj}a^+e^{-i(\Omega-\omega_{kj})t}),
\end{equation}
where aside from $u_{k1}(\lambda)$ and $v_{k1}(\lambda)$ defined in Eq.(11), we present $u_{k2}(\lambda)=u_{k2}$ and $v_{k2}(\lambda)=v_{k2}$.
Then for sufficiently long time, the
Integral $\int_0^t ds e^{i(\Omega-\omega_{kj})(t-s+i0)}$ is replaced by $\int_0^\infty ds e^{i(\Omega-\omega_{kj})(t-s+i0)}=i\frac{{\rm P}}{\Omega-\omega_{kj}}+\pi\delta(\Omega-\omega_{kj})$, where ${\rm P}$ and $\delta(x)$ denote the principal value and Dirac-delta.  
It is also remarked that in the weak coupling limit, $\sigma(t)$ in the Eq.(\ref{eqmotion}) is replaced by ${\rm Tr_r}\sigma(t)\otimes \rho_r$.
By taking the trace over the reservoir variables, and calculating the expectation values with respect to the number state $\{|0\rangle,|1\rangle\}$, one obtains the quantum master equation for ${\rm Tr}\langle n|\sigma(t)|n\rangle$ as
\begin{equation}
\left(
\begin{array}{c}
{\rm Tr_r}\dot{\sigma}_{11}(t) \\
{\rm Tr_r}\dot{\sigma}_{00}(t)
\end{array}
\right) 
=
\left(
\begin{array}{cc}
-k_{11} & k_{10} \\
k_{01} & -k_{00}
\end{array}
\right)
\left(
\begin{array}{c}
{\rm Tr_r}\sigma_{11}(t) \\
{\rm Tr_r}\sigma_{00}(t)
\end{array}
\right), \label{precession}
\end{equation}
where the transition rates are given as
\begin{eqnarray}
&&k_{11}=A_1\frac{1}{1+\frac{\lambda^2}{4}}\eta(\beta_1(\hbar\Omega-\mu_1))+A_2\eta(\beta_2(\hbar\Omega-\mu_2)) \nonumber \\
&&k_{10}=A_1\left(\frac{1}{1+\frac{i\lambda}{2}}\right)^2\eta(\beta_1(\hbar\Omega-\mu_1))+A_2\eta(\beta_2(\hbar\Omega-\mu_2)) \nonumber \\
&&k_{01}=A_1\left(\frac{1}{1-\frac{i\lambda}{2}}\right)^2(1-\eta(\beta_1(\hbar\Omega-\mu_1)))+A_2(1-\eta(\beta_2(\hbar\Omega-\mu_2))) \nonumber \\ 
&&k_{00}=A_1\frac{1}{1+\frac{\lambda^2}{4}}(1-\eta(\beta_1(\hbar\Omega-\mu_1)))+A_2(1-\eta(\beta_2(\hbar\Omega-\mu_2))). \label{quantum}
\end{eqnarray}
Here we abbreviated as $A_j=\int dk 2\pi|u_{kj}|^2\delta(\Omega-\omega_{kj})$.

Since the interaction with the probe $\lambda$ should be sufficiently small to avoid the effect of measurements, 
the coefficients of quantum master equation (\ref{quantum}) is well-approximated as 
\begin{eqnarray}
&&\frac{1}{1+\frac{\lambda^2}{4}}\cong 1 \nonumber \\
&&\left(\frac{1}{1\pm\frac{i\lambda}{2}}\right)^2\cong e^{\mp i\lambda}. \label{approx}
\end{eqnarray}
With this approximation by exponential $e^{\pm i\lambda}$, it becomes now clear that the off-diagonal element of the density matrix (\ref{spin}) indeed counts the number of electrons, i.e. the coefficient of the phase factor $e^{i\lambda m}$ of Eqs.(\ref{quantum},\ref{approx}) gives the probability $P(t,m)$ that the net number of electrons moving from the left reservoir within time $t$ is just $m$:
\begin{equation}
\sum_{j=1}^2{\rm Tr_r}\sigma_{jj}(t)=\sum_m P(t,m) e^{i\lambda m}. \label{probability}
\end{equation}
Also, the present scheme shows a mechanism to include the counting terms $e^{\pm i\lambda}$ to the quantum master equation, which has been just assumed\cite{Esposito,Esposito2,Esposito3,Utsumi2,Lebowitz}.
\section{Fluctuation theorem}
In order to further make clear that Eqs.(\ref{spin},\ref{precession}) indeed provide the electron counting statistics, 
we derive the universal symmetry of fluctuation theorem\cite{Esposito,Esposito2,Esposito3,Utsumi,Harbola,Monnai,Andrieux,Utsumi2,Lebowitz}.
This symmetry contains several out of equilibrium relations such as linear and non linear response theory\cite{Andrieux,Lebowitz}.
Since we have successfully derived the quantum master equation with counting terms, the fluctuation theorem is derived in 
a similar way as Ref.\cite{Esposito}.
A different point is the presence of the imaginary unit in the exponent of Eq.(\ref{approx}).
Let us calculate the eigenvalues of the matrix 
\begin{equation}
\left(
\begin{array}{cc}
-k_{11} & k_{10} \\
k_{01} & -k_{00}
\end{array}
\right), \label{mat}
\end{equation}
where we have assumed the first order approximation (\ref{approx})
\begin{eqnarray}
&&k_{11}=A_1\eta(\beta_1(\hbar\Omega-\mu_1))+A_2\eta(\beta_2(\hbar\Omega-\mu_2)) \nonumber \\
&&k_{10}=A_1 e^{-i\lambda}\eta(\beta_1(\hbar\Omega-\mu_1))+A_2\eta(\beta_2(\hbar\Omega-\mu_2)) \nonumber \\
&&k_{01}=A_1 e^{i\lambda}(1-\eta(\beta_1(\hbar\Omega-\mu_1)))+A_2(1-\eta(\beta_2(\hbar\Omega-\mu_2))) \nonumber \\ 
&&k_{00}=A_1(1-\eta(\beta_1(\hbar\Omega-\mu_1)))+A_2(1-\eta(\beta_2(\hbar\Omega-\mu_2))). \label{approx-2}
\end{eqnarray}
Then the eigenvalues of the matrix (\ref{mat}) is evaluated as
\begin{eqnarray}
&&g_{\pm}(i\lambda) \nonumber \\
&=&-\frac{1}{2}(A_1+A_2)\pm\sqrt{\left(\frac{A_1+A_2}{2}\right)^2-A_1A_2\left((1-e^{i\lambda})(1-\eta_1)\eta_2+(1-e^{-i\lambda})\eta_1(1-\eta_2)\right)},
\end{eqnarray}
where we have abbreviated $\eta_j=\eta(\beta_j(\hbar\Omega-\mu_j))$.
The eigenvalues satisfy a symmetry
\begin{equation}
g_{\pm}(\beta_1(\hbar\Omega-\mu_1)-\beta_2(\hbar\Omega-\mu_2)-i\lambda)=g_{\pm}(i\lambda). \label{sym}
\end{equation}

Let us derive the fluctuation theorem for the probability $P(t,m)$ of the electron transfer in the long time regime.
The Fourier series expansion of the probability $P(t,m)$ in Eq.(\ref{probability}) is expressed by a linear combination of the normal modes
\begin{equation}
\sum_m P(t,m)e^{i\lambda m}=\alpha_1e^{g_+(i\lambda)t}+\alpha_2e^{g_-(i\lambda)t}, \label{probability1}
\end{equation}
where the coefficients are given as
\begin{eqnarray}
&&\alpha_{1(2)} \nonumber \\
&&=c_{1(2)}\left(\frac{1}{2}+\frac{-(+)A_1(2\eta_1-1)+A_2(2\eta_2-1)}{4\sqrt{\left(\frac{A_1+A_2}{2}\right)^2-A_1A_2\left((1-e^{i\lambda})(1-\eta_1)\eta_2+(1-e^{-i\lambda})\eta_1(1-\eta_2)\right)}}\right), 
\end{eqnarray}
where the constants $c_{1,2}$ are determined from the initial state. 

Without loss of generality, suppose that ${\rm Re}g_+(i\lambda)>{\rm Re}g_-(i\lambda)$.
Then Eq.(\ref{probability1}) amounts to
\begin{eqnarray}
&&\lim_{t\rightarrow\infty}\frac{1}{t}\log\sum_m P(t,m)e^{i\lambda m}=g_+(i\lambda) \nonumber \\
&=&g_+(\beta_1(\hbar\Omega-\mu_1)-\beta_2(\hbar\Omega-\mu_2)-i\lambda) \nonumber \\
&=&\lim_{t\rightarrow\infty}\frac{1}{t}\log\sum_m P(t,m)e^{(\beta_1(\hbar\Omega-\mu_1)-\beta_2(\hbar\Omega-\mu_2)-i\lambda)m}, \label{log}
\end{eqnarray} 
where Eq.(\ref{sym}) is used in the second line.
Thus, we have for the finite time
\begin{eqnarray}
&&\sum_m P(t,m)e^{i\lambda m}=C_1(\lambda,t)e^{g_+(i\lambda)t} \nonumber \\
&&\sum_m P(t,m)e^{(\beta_1(\hbar\Omega-\mu_1)-\beta_2(\hbar\Omega-\mu_2)-i\lambda)m}=C_2(\lambda,t)e^{g_+(i\lambda)t}, \label{finite}
\end{eqnarray}
where the amplitudes $C_i(\lambda,t)$ behave slower than exponential $e^{g_+(i\lambda)t}$ as a function of time.
By inverse transformations of Eq.(\ref{finite}), 
we have the fluctuation theorem for the net electron transfer in the long time regime
\begin{equation}
\lim_{t\rightarrow\infty}\log\frac{P(t,m)}{P(t,-m)}=m\left(\beta_2(\hbar\Omega-\mu_2)-\beta_1(\hbar\Omega-\mu_1)\right),
\end{equation} 
which states that the ratio of the probabilities $P(t,\pm m)$ for the net number of electrons flowing through the probe   
should be balanced only by the thermodynamic affinity\cite{Andrieux,Andrieux2}.
\section{Summary}
As a model of junction systems, we considered the electron transport between the subsystem, and two reservoirs.
The electron current is monitored by the magnetic moment interaction with a localized spin probe.  
The total Hamiltonian is determined self consistently, since the total Hamiltonian includes current which depends on the Hamiltonian.        
Then we derive the quantum master equation with counting fields which describes the dynamics of the spin precession.
This procedure gives a microscopic justification of the quantum master equation with counting fields, i.e. kinetic description of the electron transport. 
As an application, we have verified the fluctuation theorem for the electron transfer. 
 
\section{Acknowledgement}
The author is grateful to Prof. S.Tasaki, Prof. P.Gaspard for fruitful discussions.
This work is supported by JSPS research fellowship for young scientists.

\end{document}